# Soliton dynamics in the Gardner–Ostrovsky equation

**R. Fariello[1], M.S. Soares[2], Y.A. Stepanyants[3,*)]**

[1]Department of Computer Science, Universidade Estadual de Montes Claros, Avenida Rui Braga, sn, Vila Mauricéia, Montes Claros, 39401-089, Minas Gerais, Brazil.
[2]Graduate Program in Computer Modeling and Systems, Universidade Estadual de Montes Claros, Avenida Rui Braga, sn, Vila Mauricéia, Montes Claros, 39401-089, Minas Gerais, Brazil.
[3]School of Mathematics, Physics and Computing, University of Southern Queensland, 487–535 West St., Toowoomba, QLD, 4350, Australia.

## Abstract

We investigate soliton formation and interactions in the non-integrable Gardner–Ostrovsky equation (alias the rotation-modified Gardner equation) with anomalous (positive) dispersion. This equation is relevant to the description of wave dynamics in plasmas and other dispersive media. We show that a special class of solitons, characterized by zero total mass and non-monotonic asymptotics, can emerge from pulse-like initial perturbations. Depending on the initial conditions, these solitons may form regular amplitude-ordered trains, irregular nonstationary ensembles of interacting localized structures, or stationary propagating multi-soliton complexes. We further demonstrate that soliton interactions in the Gardner–Ostrovsky equation are inherently inelastic, leading to the emergence of a dominant "soliton-champion" in closed systems. For example, under periodic boundary conditions, only the largest-amplitude soliton ultimately survives, while all smaller solitons are gradually eliminated through successive interactions. Despite their inelastic interactions, Gardner–Ostrovsky solitons exhibit remarkable robustness.

*) The corresponding author is Y.A. Stepanyants, email: Yury.Stepanyants@unisq.edu.au.

## 1. Introduction

The role of the Gardner equation in the description of long nonlinear internal waves in the ocean is well known (see, for example, [1, 2]). This equation incorporates both quadratic and cubic nonlinearities, the latter becoming particularly important when the coefficient of the quadratic nonlinearity becomes anomalously small. The Gardner equation can also serve as a model for the description of surface water waves and waves in other media. It is a completely integrable equation possessing soliton solutions, an infinite number of conservation laws, and many other remarkable mathematical properties (see, for example, [3, 4]).

When the medium is subject to rotation, an additional dispersive term arises in the equation. This effect was first demonstrated by Ostrovsky [5] for internal waves in a shallow ocean and later by Grimshaw [6] for internal waves in a deep ocean. Incorporating cubic nonlinearity and rotational (Coriolis) dispersion into the classical Ostrovsky equation leads to the generalized Gardner–Ostrovsky (GO) equation:

$$\frac{\partial}{\partial x}\left(\frac{\partial \eta}{\partial t} + c\frac{\partial \eta}{\partial x} + \alpha\eta\frac{\partial \eta}{\partial x} + \alpha_1\eta^2\frac{\partial \eta}{\partial x} - \beta\frac{\partial^3 \eta}{\partial x^3}\right) = \gamma\eta, \tag{1.1}$$

where function $\eta(x, t)$ in the oceanic context denotes a wave perturbation of a pycnocline in a density stratified ocean, $c$ is the speed of long linear waves, and $\alpha$, $\alpha_1$, $\beta$, and $\gamma$ are coefficients that are determined by the bathymetry and stratification. The coefficient $\gamma$ is usually positive, whereas other coefficients, $\alpha$, $\alpha_1$, and $\beta$, can be of either sign [7]. Such an equation has been used in many papers in application to oceanic waves, see, for example [7–10, 2]. However, it can be used also for the description of waves of other types such as waves in plasma, solids, and other media. This equation is non-integrable and does not have soliton solutions in media with negative dispersion when $\beta < 0$. The initial pulse in the form of a Gardner soliton (i.e. an exact solution of Eq. (1.1) with $\gamma = 0$) gradually decays emitting quasi-linear waves; this process was studied in [11].

The situation is different when $\beta > 0$. In this case, Eq. (1.1) may admit soliton solutions. However, the properties of the GO equation (1.1) for $\beta > 0$ have not been investigated so far, and even the existence of solitary-wave solutions has remained unexplored. In this paper, we numerically construct solitary-wave solutions, which we refer to as solitons for brevity. We demonstrate the existence of a family of such solutions that can form bound states, giving rise to multi-soliton structures. Numerical simulations show that these solitons are robust, can emerge from a wide range of pulse-like initial perturbations, and remain stable during mutual interactions. In a closed system with periodic boundary conditions, soliton interactions lead to the formation of a "soliton champion", namely the soliton with the largest amplitude, which eventually eliminates smaller-amplitude solitons (see [12] and references therein). Similar phenomena with soliton-champion formation were observed in recent publications within various equations [13–17], however, in each case some specific features were observed.

The remainder of the paper is organised as follows. Section 2 discusses the linear properties of the GO equation and presents three conserved quantities. Section 3 introduces a normalised form of the GO equation and describes the numerical construction of its solitary-wave solutions. Section 4 demonstrates that solitons in the GO equation can emerge from initial pulses with zero total mass. Section 5 presents numerical results on soliton interactions and shows that, in a closed system with periodic boundary conditions, a soliton-champion can develop from the soliton with the largest amplitude. Finally, Section 6 summarises the main results of the study.

**2. Linear properties of the Gardner–Ostrovsky equation**

Considering waves of small amplitudes and neglecting the nonlinear terms in Eq. (1.1), we derive for an infinitesimal perturbation $\eta \sim \exp[i(\omega t - kx)]$ the dispersion relation:

$$\omega = ck + \beta k^3 + \frac{\gamma}{k}, \tag{1.2}$$

where $\omega$ is a wave frequency and $k$ is a wavenumber. From here, we obtain the expressions for the phase and group velocities (for the unidirectional wave propagation, we assume that $k > 0$):

$$V_{ph} \equiv \frac{\omega}{k} = c + \beta k^2 + \frac{\gamma}{k^2}; \qquad V_g \equiv \frac{d\omega}{dk} = c + 3\beta k^2 - \frac{\gamma}{k^2}. \tag{1.3}$$

The dispersion relation $\omega(k)$ is shown in Fig. 1, and the dependences of phase and group velocities on the wavenumber $k$ are given in Fig. 2.

As follows from Eq. (1.3) and Fig. 2a, when $\beta < 0$, the phase velocity of linear perturbations is unlimited (see blue line 2 in Fig. 2a), whereas the group velocity cannot exceed $c$ (see Fig. 2b). The maximum group velocity is attained at $k_g = \sqrt[4]{-\gamma/3\beta}$. The situation is reverse when $\beta > 0$ where the phase velocity of linear perturbations is restricted from below and it cannot be less than $c$ (see Fig. 2a); whereas the group velocity is unlimited (see Fig. 2b). The minimum phase velocity occurs at $k_p = \sqrt[4]{\gamma/\beta}$ and equals $V_{min} = c + 2\,(\beta\gamma)^{1/2}$. Such a dependence of the phase velocity on the wave number leads to an important physical consequence that stationary solitary waves may exist only when $\beta > 0$, whereas in the case of $\beta < 0$ their existence is impossible (cf. the similar situation in the Ostrovsky and rotation modified Benjamin–Ono equations [18, 14]).

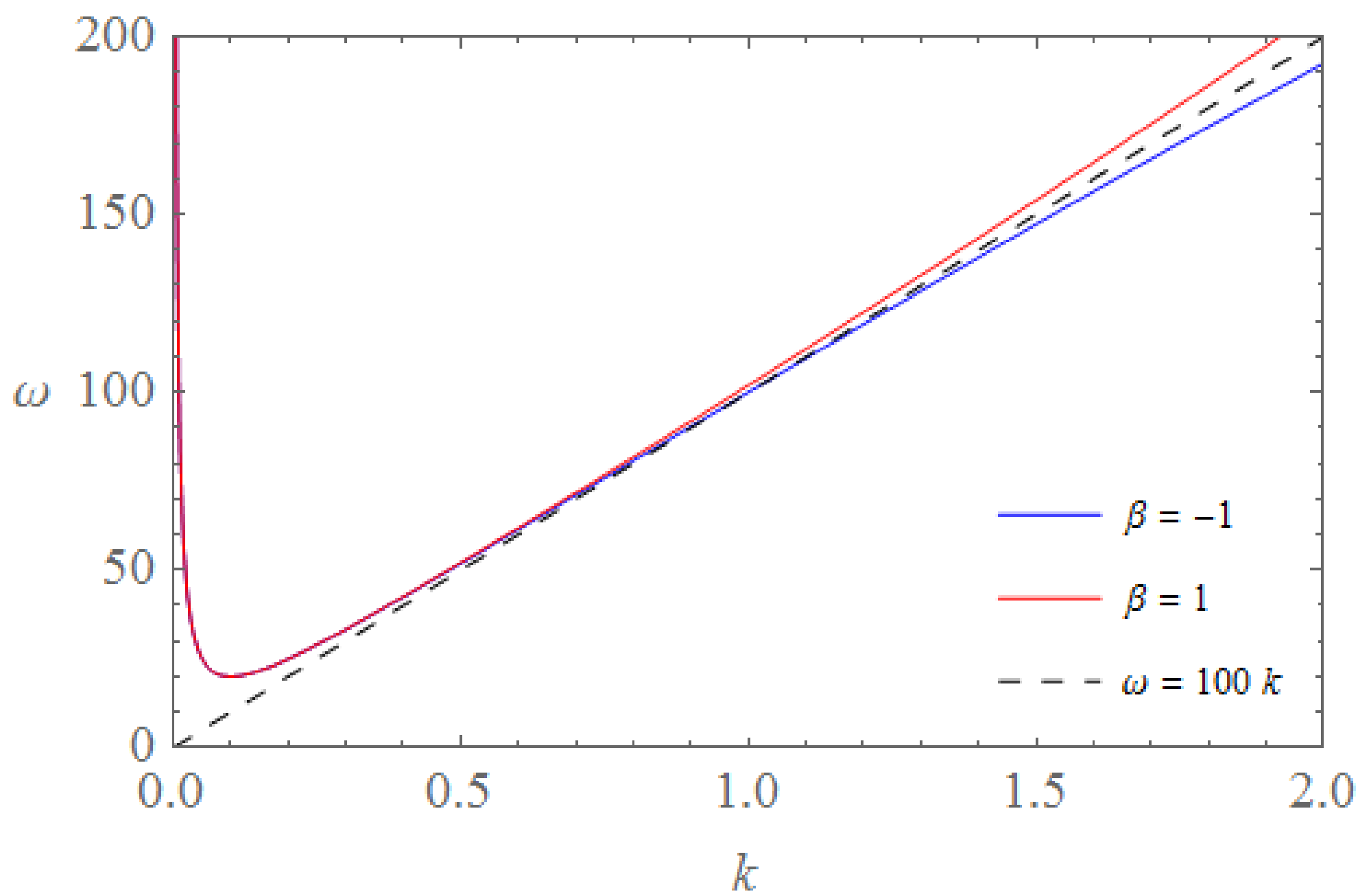

Fig. 1 (colour online). The dispersion dependence of the wave frequency on the wave number for the normalised GO equation (1.1) with $c = 100$ and $\gamma = 1$. Line 1 pertains to the positive ("anomalous") dispersion case with $\beta = 1$, and line 2 – to the negative ("normal") dispersion case with $\beta = -1$. The diagonal dashed line shows the dispersionless dependence $\omega = ck$, where $c = 100$.

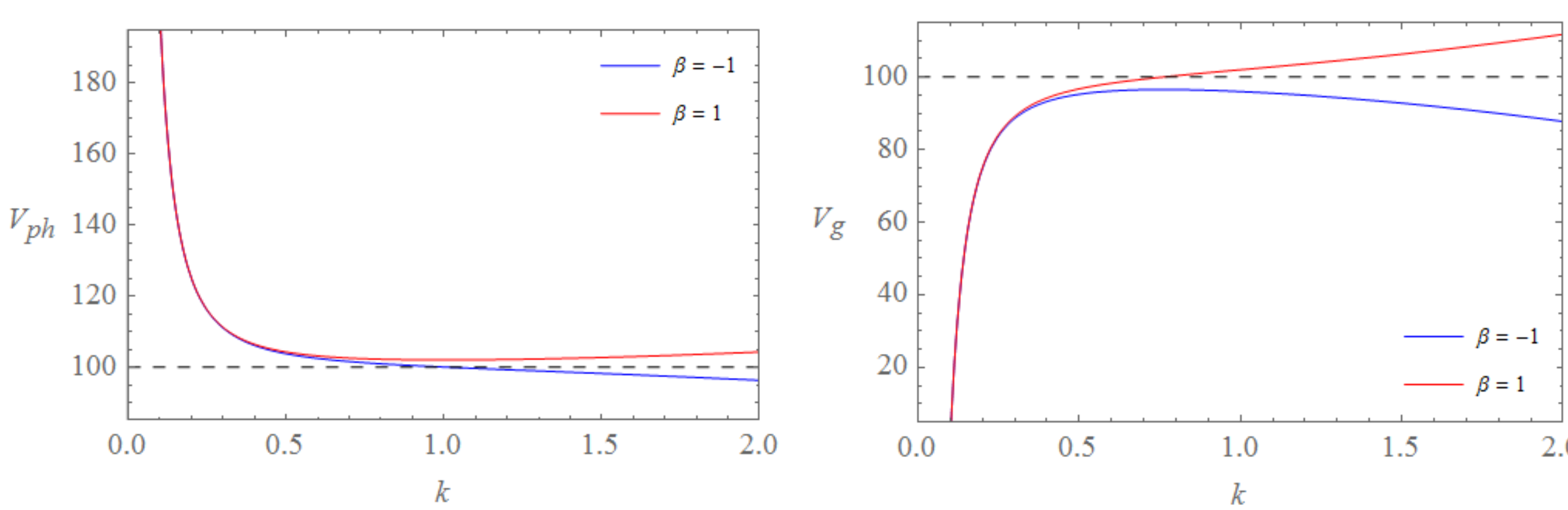

a) b)

Fig. 2 (colour online). Dependences of the phase (left panel) and group (right panel) velocities on the wave number for different signs of the small-scale dispersion $\beta$. Lines 1 pertain to the positive dispersion case with $\beta = 1$ and lines 2 – to the negative dispersion case with $\beta = -1$. Horizontal dashed lines show the velocity $c = 100$.

The GO equation complies with the following conserved quantities:

- $I_1 \equiv \int \eta(x,t)\, dx = 0$ – "mass" conservation. (1.4)

- $I_2 \equiv \frac{1}{2}\int \eta^2(x,t)\, dx = \text{const}$ . – "momentum" conservation. (1.5)

- The Hamiltonian: $I_3 \equiv \frac{1}{2}\int \left[ -\beta(\eta_x)^2 + c\eta^2 + \frac{1}{3}\alpha\eta^3 + \frac{1}{6}\alpha_1\eta^4 + \gamma\upsilon^2 \right] dx = \text{const.}, \quad \eta = \upsilon_x$ . (1.6)

In these formulae, integration is assumed along the whole $x$-axis for localized perturbations or over the period for periodic perturbations. The constants in Eqs. (1.5) and (1.6), are determined by the initial conditions, whereas the mass conservation integral is essentially the limitation on the class of admissible perturbations, which requires a zero mean value of perturbation per period or zero total "mass" of the localized perturbation. This limitation naturally follows from the condition of applicability of the GO equation.

### 3. Solitary waves in the Gardner-Ostrovsky equation with positive dispersion

The existence of solitary wave solutions in the GO equation can be expected only in the case of anomalous dispersion when $\beta\gamma > 0$ (in what follows, we assume that $\gamma > 0$, therefore solitary waves can exist only when $\beta > 0$). In this case, phase wave speeds of small amplitude linear waves are restricted from below by $V_{min} = c + 2(\beta\gamma)^{1/2}$; therefore, a source (e.g., soliton) can move with a constant speed $V < V_{min}$ without a resonance with linear waves and energy losses on their generation. Such situations apparently can occur in magnetized rotating plasma and in some other physical media with positive dispersion. To our best knowledge, solutions of the GO equation in the form of solitary waves with the zero total mass were not studied yet. Therefore, using the Petviashvili method [19, 20], we construct numerically a family of solitons as functions of their speeds.

Let us present first Eq. (1.1) in the dimensionless form:

$$\frac{\partial}{\partial \xi}\left( \frac{\partial u}{\partial \tau} - u\frac{\partial u}{\partial \xi} + \delta u^2 \frac{\partial u}{\partial \xi} - \frac{\partial^3 u}{\partial \xi^3} \right) = u. \tag{2.1}$$

This can be done with the help of the transformation:

$$\tau = t\gamma\left(\frac{\beta}{\gamma}\right)^{1/4}, \quad \xi = \left(\frac{\gamma}{\beta}\right)^{1/4}(x - ct), \quad u = -\eta\frac{\alpha}{(\beta\gamma)^{1/2}}, \quad \delta = \frac{\alpha_1}{\alpha^2}(\beta\gamma)^{1/2}, \tag{2.2}$$

where $\delta$ is an arbitrary parameter which can be of either sign. Further we will study solutions to Eq. (2.1) with different values of $\delta$. Searching for a stationary solution, we assume that function $u(\xi, \tau)$ depends on one variable only $z = \xi - V\tau$, where $V$ is a dimensionless wave speed. Then, Eq. (2.1) reduces to the ODE:

$$\frac{d^2}{dz^2}\left(\frac{d^2u}{dz^2} - Vu - \frac{1}{2}u^2 + \frac{1}{3}\delta u^3\right) = u\ . \tag{2.3}$$

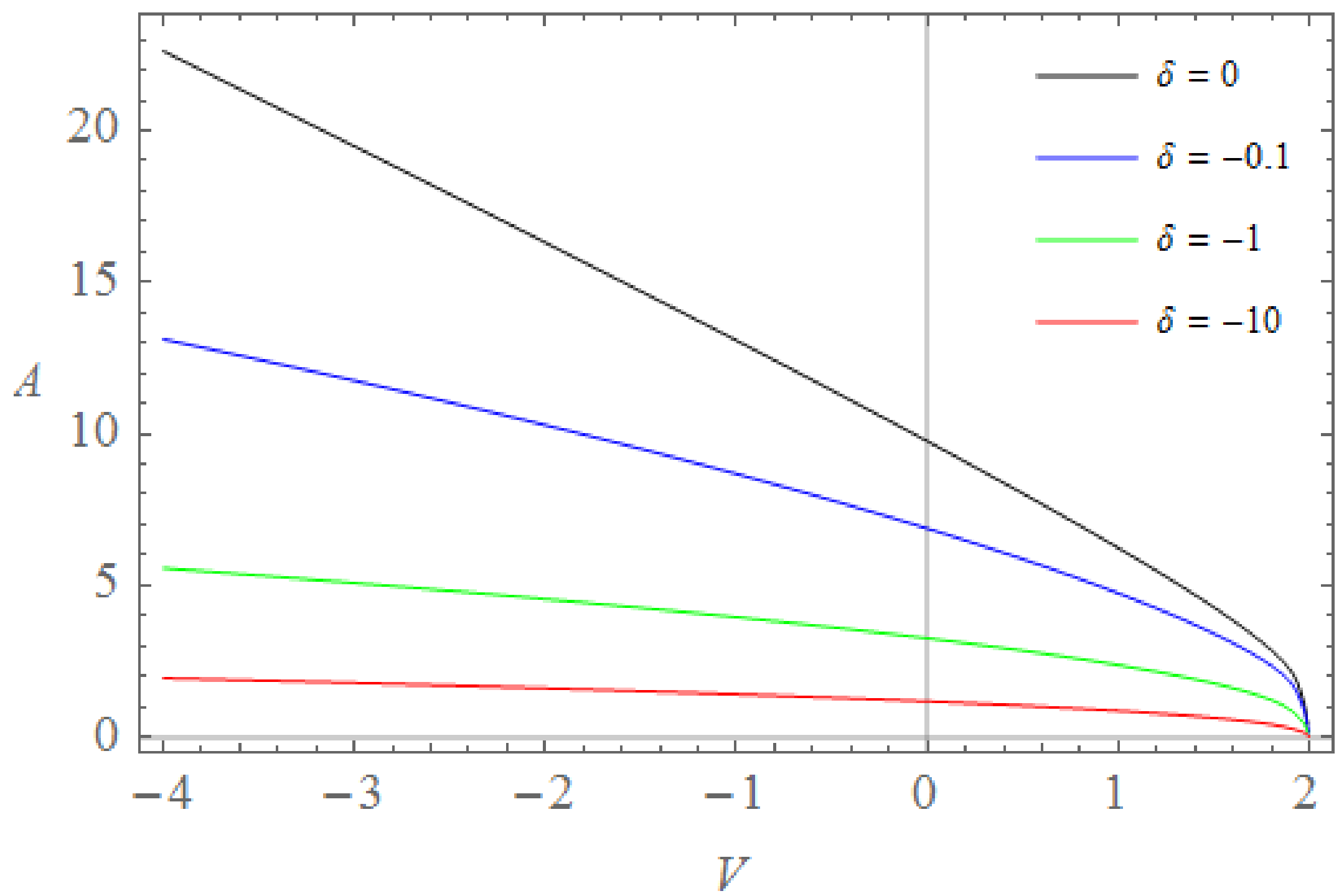

Fig. 3 (colour online). The numerically obtained dependences of soliton amplitude $A$ on speed $V$ within the dimensionless GO equation (2.1) with different coefficients $\delta$. The critical soliton speed $V_c = 2$ corresponds to the minimal phase speed within the normalised Eq. (2.1).

The shape and amplitude of a solitary wave, if it exists, depend on two parameters, the wave speed $V$ and the nonlinear cubic coefficient $\delta$. When $\delta = 0$, Eq. (2.3) reduces to the Ostrovsky equation with positive dispersion, whose soliton solutions were studied (see, for example, [21, 22, 13]). For $\delta \neq 0$, we managed to construct soliton solutions only when it is negative. Figure 3 shows the dependences of soliton amplitudes on a speed for four values of $\delta = 0$, −0.1, −1, and 10. The reference case with $\delta = 0$ corresponds to the Ostrovsky equation studied in [21, 13] within Eq. (2.1) with a different normalisation. As one can see from Fig. 3, qualitatively soliton structures are the same for any value of $\delta \le 0$, but soliton amplitudes decrease with increasing $|\delta|$ for a fixed value of $V$ and decrease when $V \to 2$ for a fixed $\delta$.

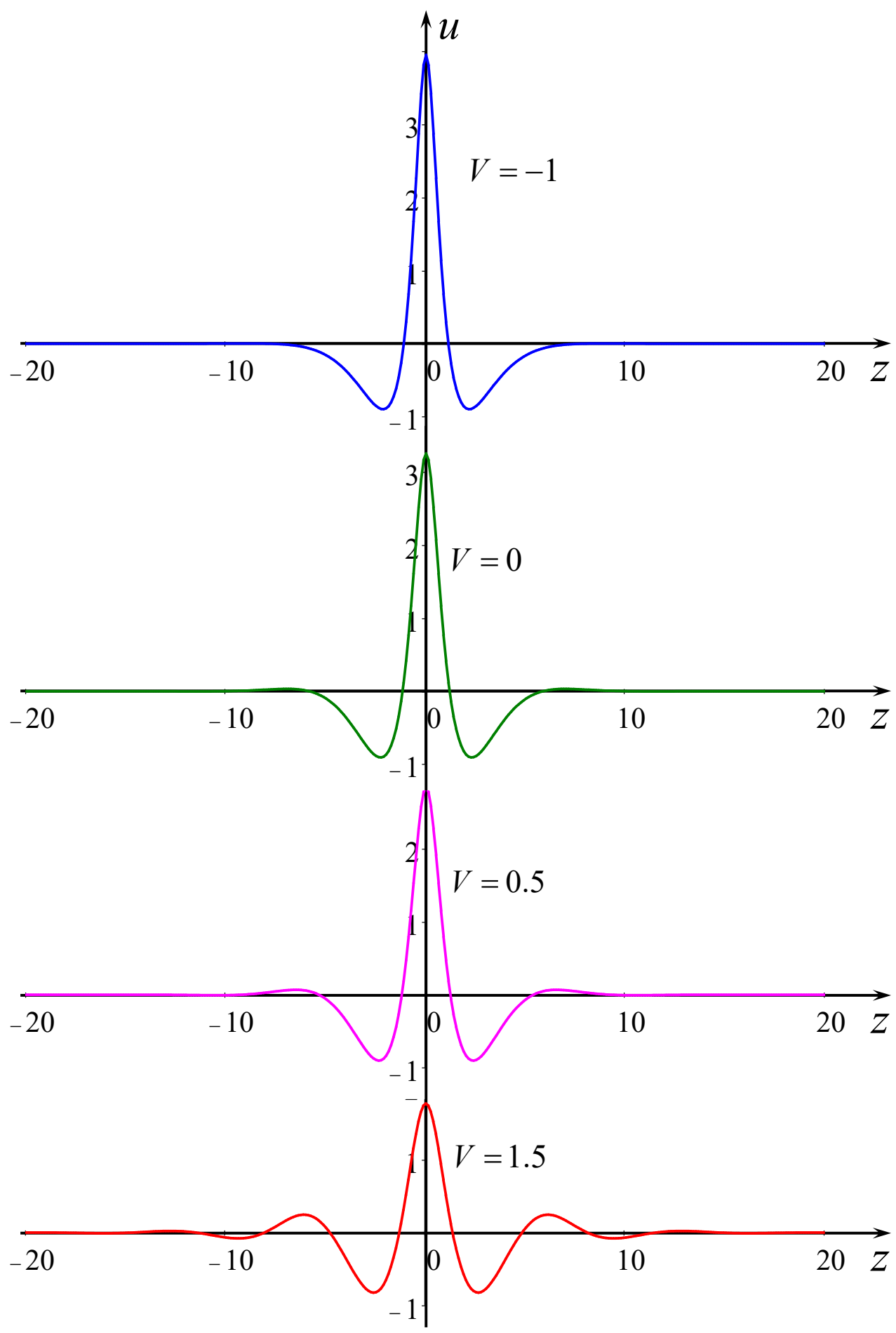

Figure 4

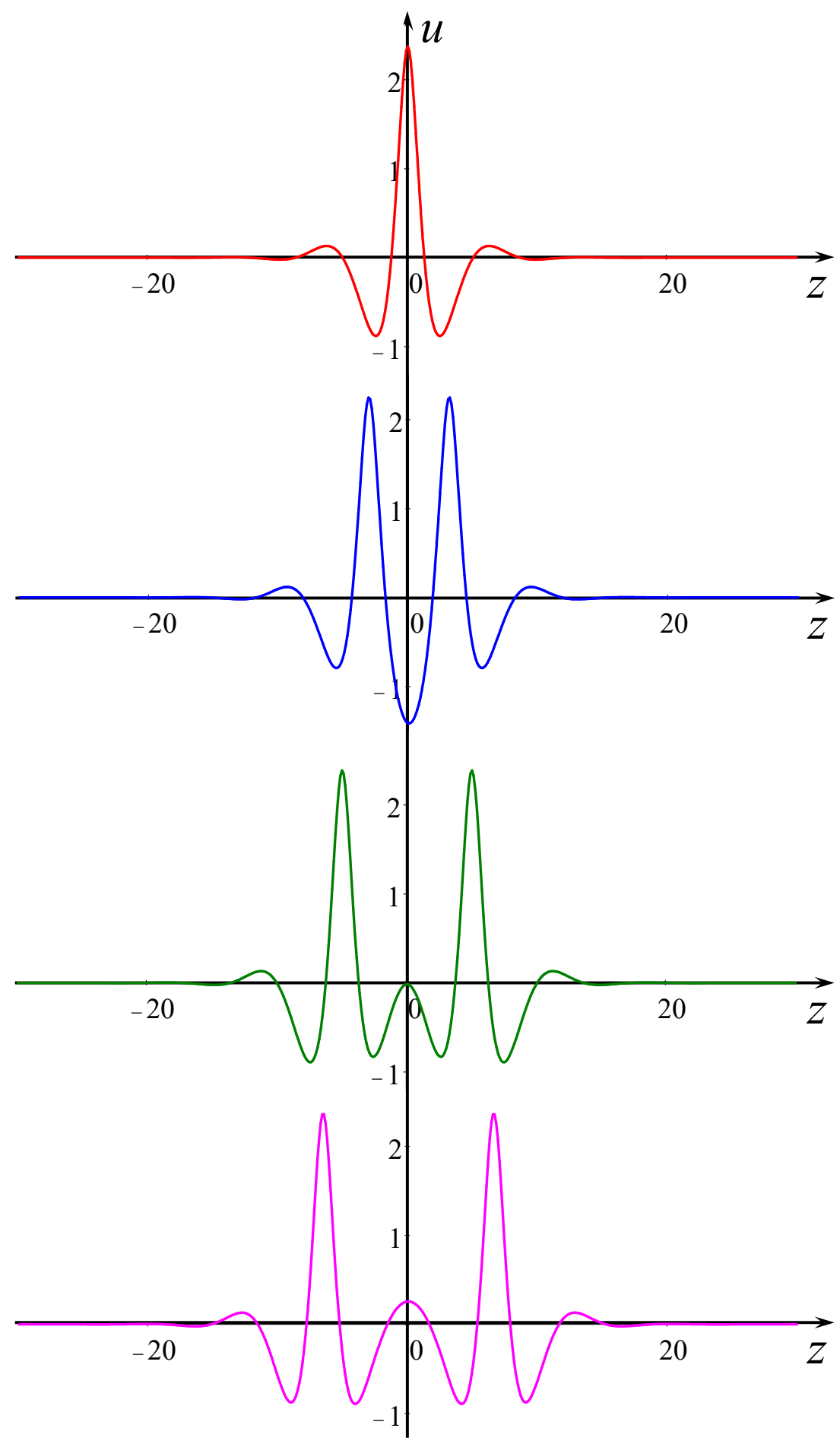

Figure 5

Fig. 4 (colour online). Soliton shapes for different speeds $V$ and fixed value of $\delta = -1$.

Fig. 5 (colour online). Bi-solitons – stationary bound states of two solitons. A single soliton with a nonmonotonic profile at $V = 1$ is shown in the upper frame; in the subsequent frames one can see a family of stationary bi-solitons with the same speed.

The dependence of soliton shape on $V$ is shown in Fig. 4 for $\delta = -1$. As the parameter $V$ approaches its upper critical value, $V_c = 2$, an increasing number of oscillations appear in the soliton profile, and the solitons progressively resemble wave trains (see, for example, frame 5 in Fig. 4). However, in such solitons, the carrier and envelope waves propagate at the same velocity, unlike the conventional envelope solitons of the nonlinear Schrödinger equation (see, e.g., [23]). For $V > V_c$, soliton solutions with zero asymptotics at infinity do not exist.

The nonmonotonic character of solitary wave profile suggests that solitons are able to form stationary bound states – bi-solitons, tri-solitons, etc., like in the Ostrovsky equation with positive dispersion [21, 13] or in the rotation-modified Benjamin–Ono equation [14]. Such stationary formations were also obtained numerically for the GO equation by means of the Petviashvili method; the examples are shown in Fig. 5 for $\delta = -1$; similar solution exist for other negative values of $\delta$. The bound states can be stable when a peak of each solitary wave coincides with a local minimum of the neighbouring soliton or unstable when the wave peaks coincide with local maxima of the neighbouring soliton. Apparently, it is possible to create even infinite stationary waves consisting of solitons and multi-solitons with random alternation of maxima and minima. Perhaps, two coupled solitons with nonmonotonic profiles can form an oscillating pair, termed a *breather*, similar to that observed in experiments with electromagnetic transmission lines described by the Kawahara-type equation [24, 25]; such breathers can emerge near stable states.

Despite the existence of stationary solitary waves in the case of the "anomalous" positive dispersion when $\beta > 0$ in Eq. (1.1), their robustness and evolutionary character, i.e. possibility to emergence from arbitrary initial perturbations is not obvious. Therefore, below we investigate the possibility of their formation from arbitrary pulse-type initial perturbations and their stability with respect to interaction with each other.

## 4. Emergence of solitons from initial pulses with zero masses

As a first step, we investigated the emergence of solitons from soliton-like pulses. To this end, a soliton with aperiodic asymptotics was constructed using the Petviashvili method for $V = 0.5$ and amplitude $A_0 = 2.85$ (see the pink curve in Fig. 4). The stationarity of this solution was verified using an independent numerical code developed specifically for solving the GO equation (1.1) and described in [26]. The computations confirmed that the solution is stationary and stable, at least with respect to numerical round-off errors.

The same solution was then used as a reference function and stretched vertically by multiplying it by a factor $F = 1.2$. During the subsequent evolution, the initial pulse emitted a small-amplitude dispersive ripple and, within a relatively short time, evolved into another stationary soliton with a slightly larger amplitude (see Fig. 6). A similar behaviour was observed when the reference profile was multiplied by the factor $F = 0.8$ (see Fig. 7). The temporal evolution of the pulse amplitudes, normalised by their initial values, is shown in Fig. 8 for both cases. The normalised amplitudes asymptotically approach 1.12 and 0.84 for $F = 1.2$. and $F = 0.8$, respectively.

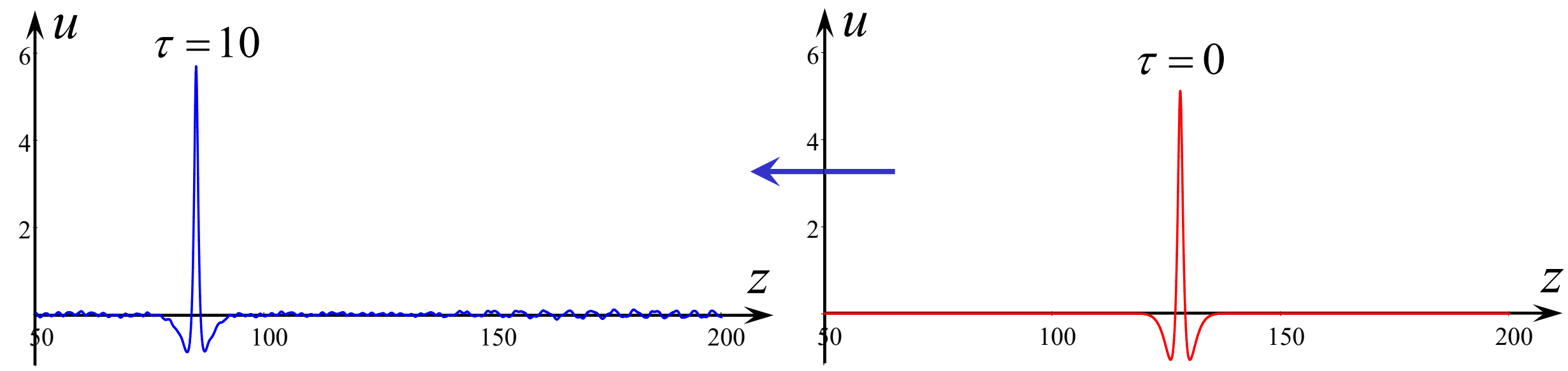

Fig. 6. Evolution of soliton-like initial pulse with the multiplication factor $F = 1.2$.

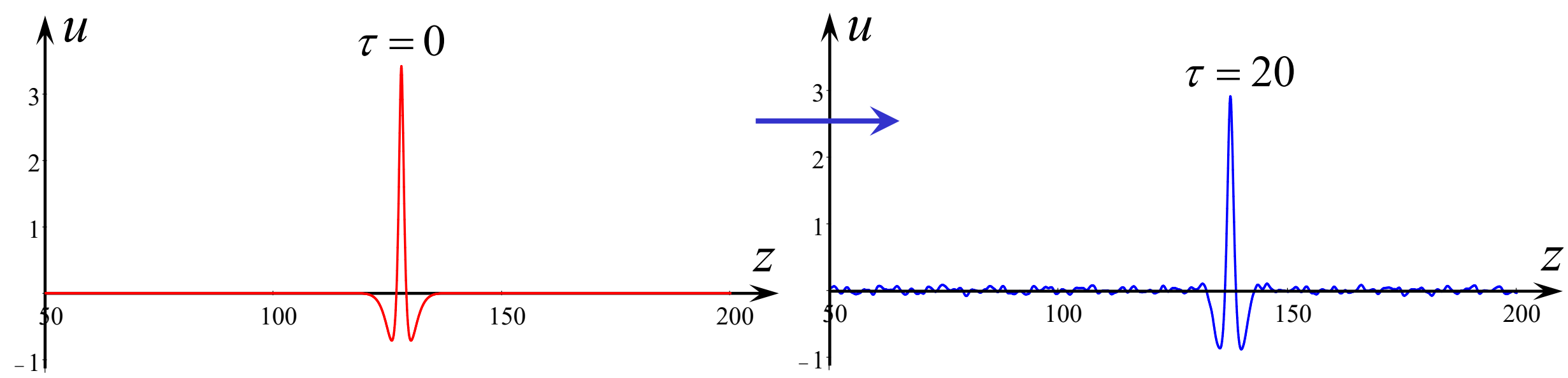

Fig. 7. Evolution of soliton-like initial pulse with the multiplication factor $F = 0.8$.

However, when the amplitude factor $F$ was sufficiently small, the initial pulse did not evolve into a soliton. Instead, it gradually transformed into a decaying oscillatory wavetrain. Figure 9 illustrates the time dependence of the pulse amplitude for $F = 0.3$. In this case, the amplitude asymptotically approaches the linear trend $A(\tau) = -0.146\tau + 0.76$, shown by the thin blue line in Fig. 9. The corresponding pulse profile at $\tau = 1.448$ is presented in the inset of that figure.

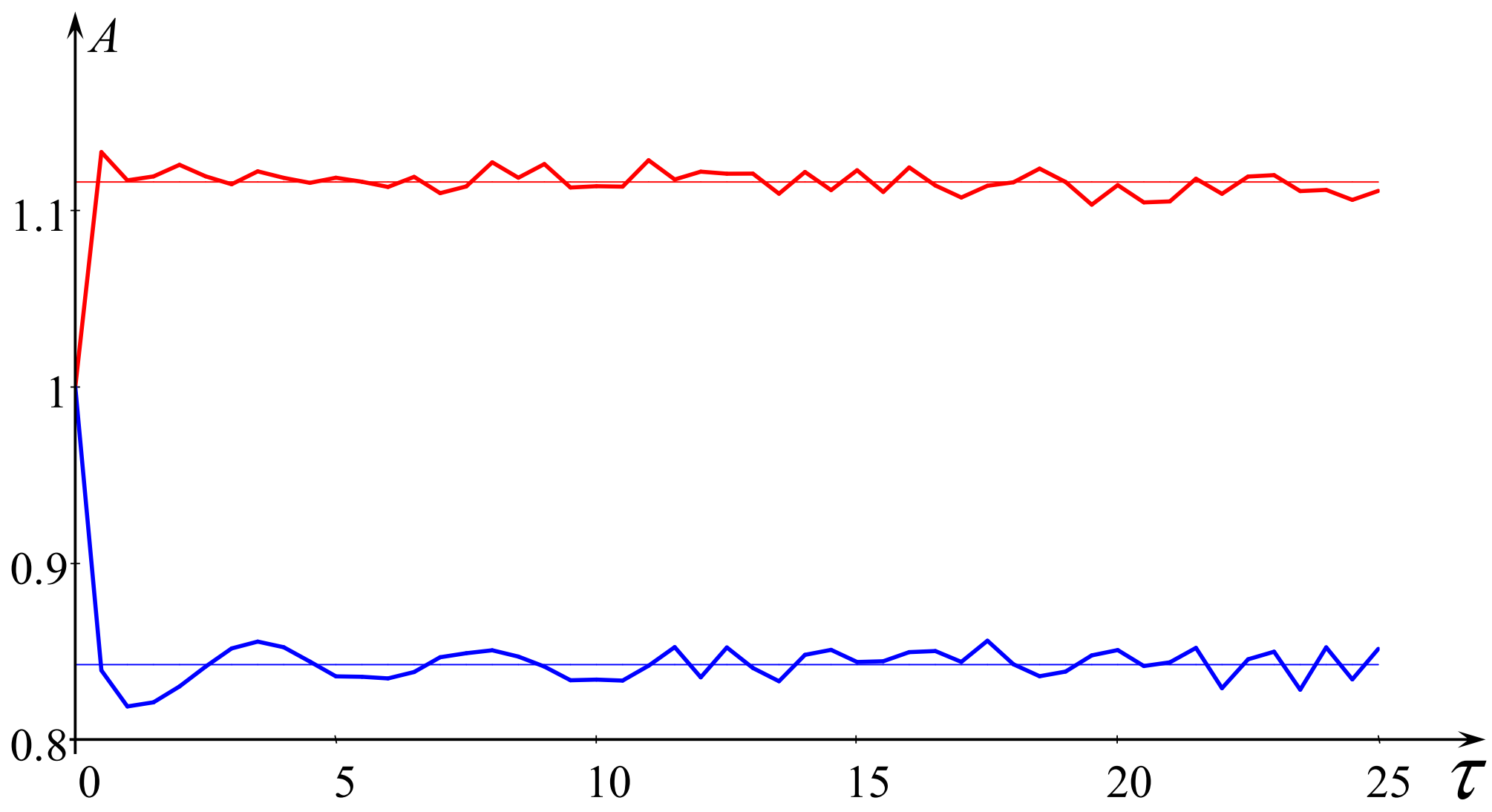

Fig. 8. Dependences of normalised pulse amplitudes in the process of soliton formations. Red line pertains to the case with the amplitude factor $F = 1.2$, blue line pertains to the case with $F = 0.8$.

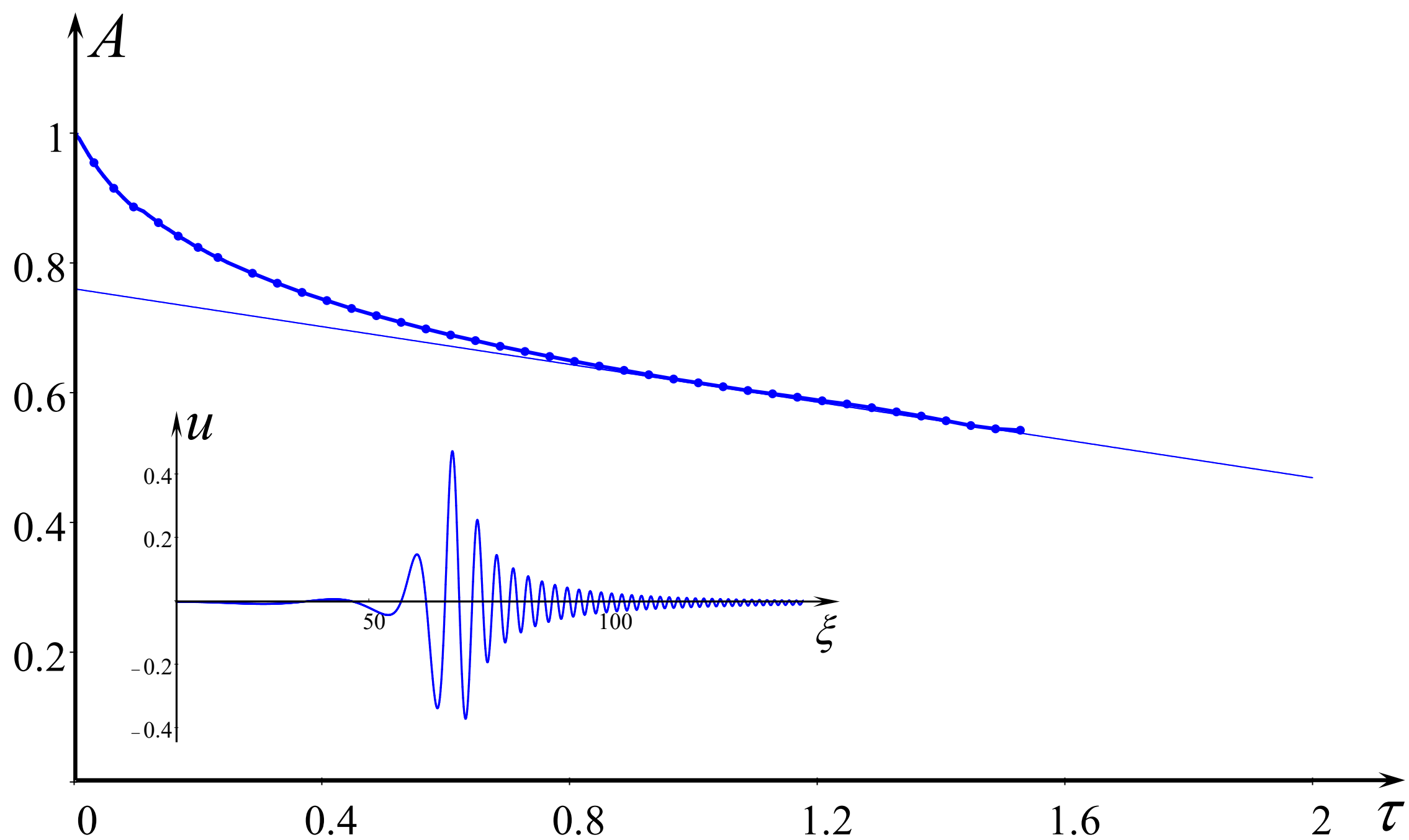

Fig. 9. The dependence of the normalised amplitude on time of a soliton-like initial pulse with the multiplier $F = 0.3$. Thin line shows the asymptotic dependence of the normalised pulse amplitude. The insertion shows the pulse shape at $\tau = 1.448$.

As follows from these results, solitons emerge only from initial pulses whose amplitudes exceed a certain threshold. We did not attempt to determine the threshold amplitude for soliton formation in the present study. For a fixed value of the parameter $\delta$, this threshold is expected to depend not only on the pulse amplitude but also on its width and shape. To illustrate this point, we rescale the original Eq. (1.1) in an alternative manner, assuming that the initial pulse is represented by a one-parameter function $f(x/D)$ of unit amplitude, such that $\eta(x,0) = Af(x/\Delta)$ where $A$ is the pulse amplitude and $\Delta$ is its characteristic width. Then, by introducing new variables

$$\theta = \frac{\alpha A}{\Delta}t, \quad \zeta = \frac{x-ct}{\Delta}, \quad \upsilon = -\frac{\eta}{A}, \tag{4.1}$$

we reduce Eq. (1.1) to the following dimensionless form:

$$\frac{\partial}{\partial\zeta}\left(\frac{\partial\upsilon}{\partial\theta} - \upsilon\frac{\partial\upsilon}{\partial\zeta} + q\upsilon^2\frac{\partial\upsilon}{\partial\zeta} - \frac{1}{\mathrm{Ur}}\frac{\partial^3\upsilon}{\partial\zeta^3}\right) = \frac{1}{\mathrm{So}}\upsilon, \tag{4.2}$$

where $\mathrm{Ur} = \alpha A\Delta^2/\beta$ is the Ursell parameter for the KdV equation (alias the similarity parameter $\sigma^2$ defined in [27]), and $q = \alpha_1 A/\alpha$ and $\mathrm{So} = \alpha A/\gamma\Delta^2$ are two other dimensionless parameters; all three parameters are independent. The first one, Ur, measures the ratio of the nonlinearity to small-scale dispersion. In the KdV equations, this parameter controls the number of emerging solitons; for a single soliton Ur = 12. Similarly, the parameter So measures the ratio of the nonlinearity to large-scale dispersion (see [28]). The last parameter, $q$, measures the influence of a cubic nonlinearity relative to quadratic nonlinearity. Therefore, the Cauchy problem for Eq. (3.1) with the normalized initial condition $\upsilon(\zeta, 0) = f(\zeta)$ contains again the same number of independent parameters, Ur, $q$, and So as Eq. (2.1) which contains independent parameters $A$, $\Delta$, and $\delta$. In the result of this, the identification of threshold surfaces in the three-dimensional parameter space that separate domains corresponding to different number of emerging solitons becomes extremely difficult in contrast with what can be done for the KdV equation [27]. For the three cases considered above where the width $\Delta$ of the initial pulses was the same, and amplitudes were different $A_1 = 3.42$, $A_2 = 2.28$, and $A_3 = 0.85$, these parameters were $\mathrm{Ur}_1 = 3.42\Delta^2$, $\mathrm{Ur}_2 = 2.28\Delta^2$, and $\mathrm{Ur}_3 = 0.85\Delta^2$; $q_1 = -3.42$, $q_2 = -2.28$, and $q_3 = -0.85$; $\mathrm{So}_1 = 3.42/\Delta^2$, $\mathrm{So}_2 = 2.28/\Delta^2$, and $\mathrm{So}_3 = 0.85/\Delta^2$. Only in the case with parameters $\mathrm{Ur}_3 = 0.85\Delta^2$, $q_3 = -0.85$, and $\mathrm{So}_3 = 0.85/\Delta^2$ soliton did not emerge.

## 5. Soliton interactions

As noted above, the GO equation is non-integrable, whereas the Gardner equation is completely integrable [2–4]. Non-integrability leads to inelastic soliton interactions; that is, after the interaction of two solitons, their parameters, including amplitude, velocity, and width, change slightly, and a small-amplitude wavetrain is typically generated. In closed systems, such as those with periodic boundary conditions, solitons can undergo multiple interactions, emitting a new wavetrain during each collision. As a result, the space between solitons gradually becomes filled with a quasi-random wave background. Moreover, small-amplitude solitons tend to lose energy during repeated interactions, transferring part of its energy to larger-amplitude solitons, which consequently grow even larger, and part to the background wave field.

This process, leading to the formation of a so-called *soliton-champion* in closed non-integrable systems, was first reported by Krylov and Yan'kov [29] and subsequently investigated in a number of studies [12–17]. A similar phenomenon may be expected in a closed system governed by the GO equation with periodic boundary conditions.

Below, we present the results of numerical simulations performed within Eq. (2.1) with $\delta = -1$ in a periodic domain of the total length $L = 256$. In the first experiment, two preliminary generated solitons of different amplitudes $A_1 = 3.954$ and $A_2 = 2.373$ with corresponding velocities $V_1 = -1$ and $V_2 = 1$ respectively (see Fig. 3), were placed in the computational domain at some distances from each other; their positions were $S_1 = 192$ and $S_2 = 64$. The initial soliton profiles are shown in the upper panel of Fig. 10 at $\tau = 0$. Having different amplitudes, solitons moved in the opposite direction with the same speeds and eventually they collided. Already after the first collision, the small-amplitude soliton was destroyed and transformed into a nonstationary wavetrain shown in the middle panel of Fig. 10. In the process of interaction, solitons emitted small-amplitude low-frequency waves propagating to the left and high-frequency waves propagating to the right. Eventually, all domain was filled by small-amplitude background wave field. The amplitude of the large soliton increased after the interaction by 8.5% as shown in Fig 11.

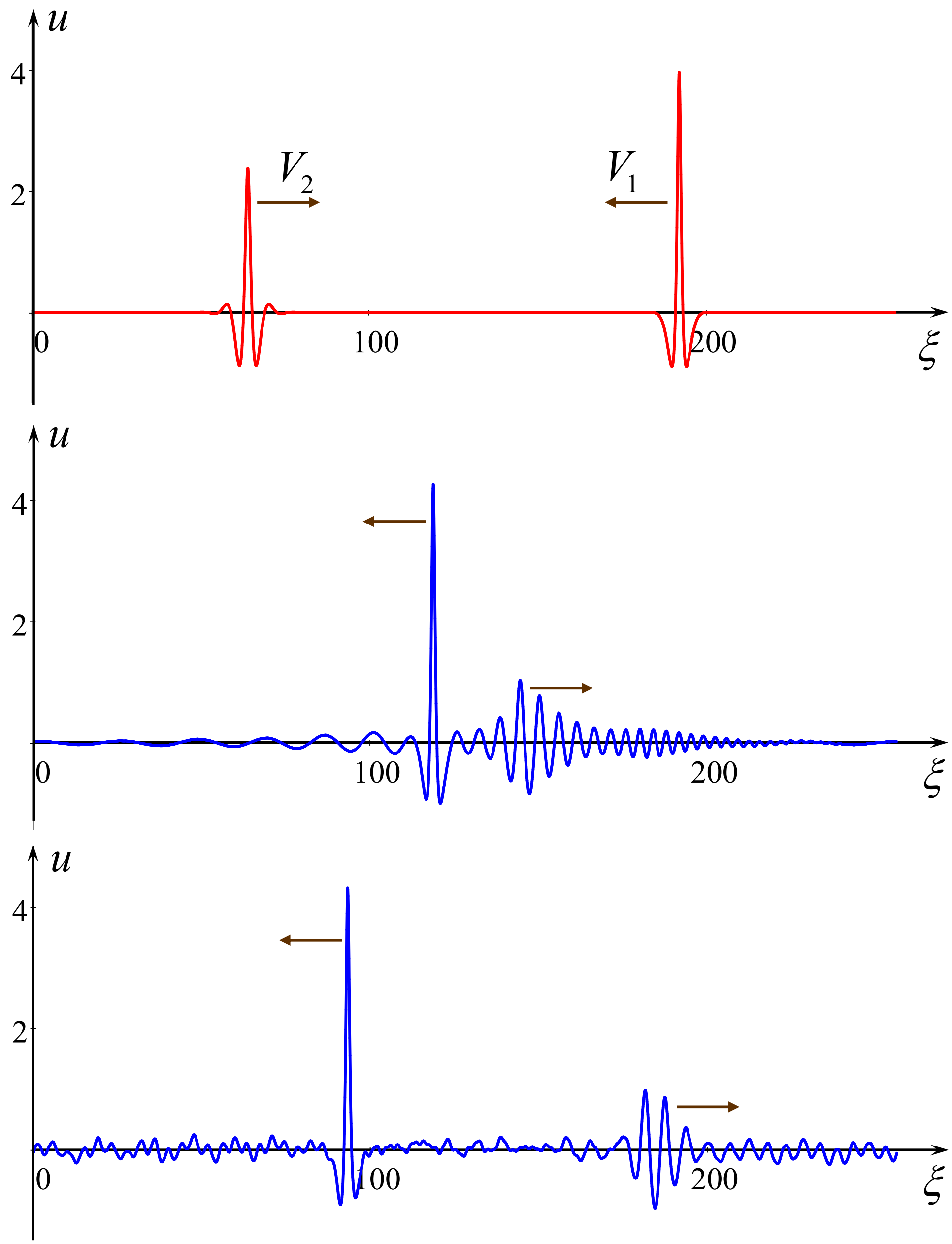

Fig. 10. Head-on collision of two solitons of different amplitudes moving initially with speeds $V_{1,2} = \pm 1$. The upper panel shows the initial time moment $\tau = 0$, the middle panel shows pulses just after the collision at $\tau = 79$, and lower panel shows pulses at $\tau = 96$.

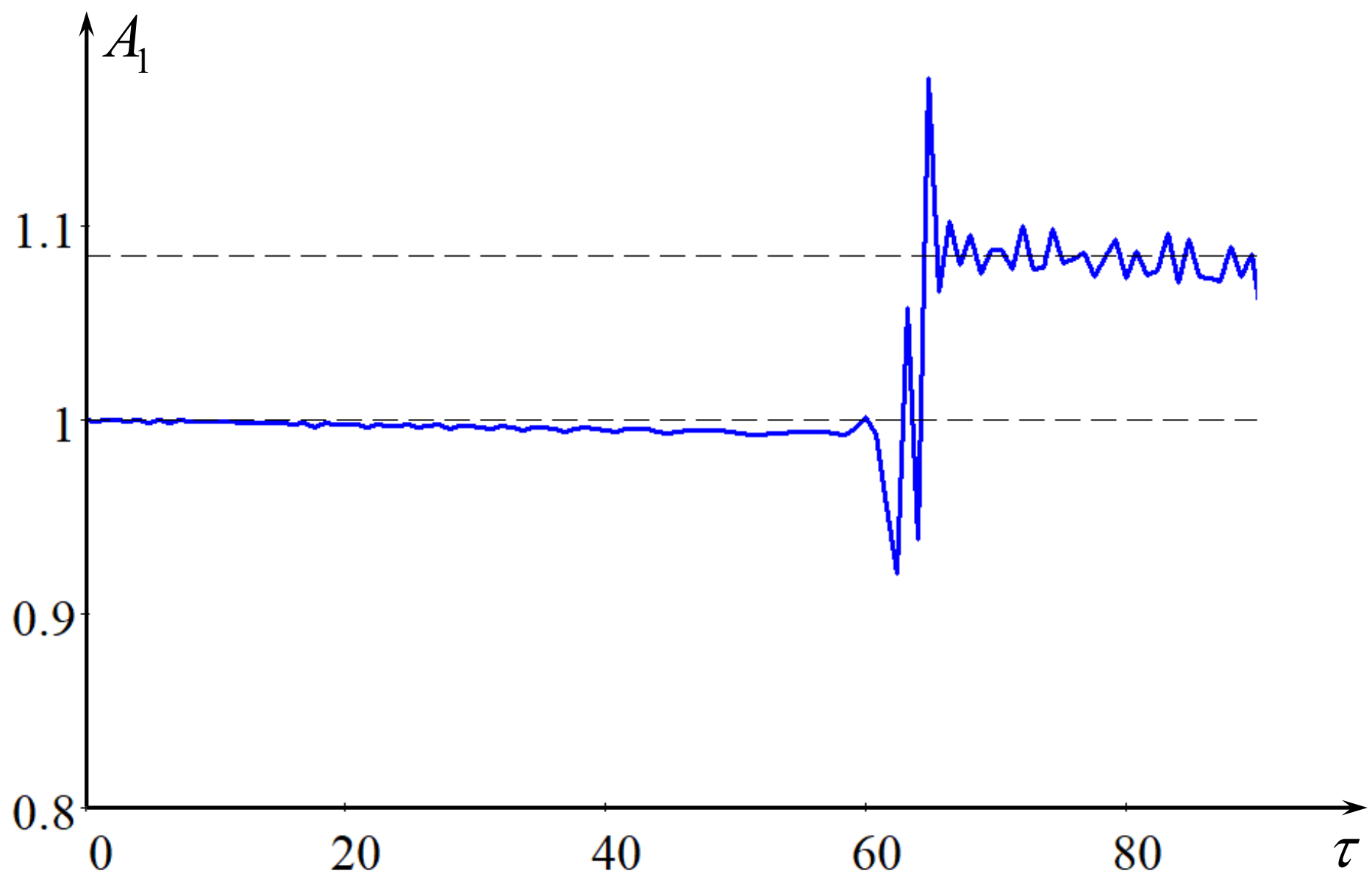

Fig. 11. Time dependence of the largest soliton amplitude.

A similar process, in which a small-amplitude bi-soliton is eventually eliminated by a large-amplitude soliton, was also observed. However, the dynamics of this process depend on the intrinsic stability of the bi-soliton, namely whether it is stable with respect to small perturbations or not (see Section 3 and the discussion associated with Fig. 5).

Figure 12 illustrates the interaction between a soliton with initial amplitude $A_1$ = 3.954, propagating to the left with velocity $V_1 = -1$, and an unstable bi-soliton with initial amplitude $A_2$ = 2.25, propagating to the right with velocity $V_2 = 1$. Owing to its instability, the bi-soliton rapidly disintegrates under the influence of small numerical round-off errors, producing a non-stationary wavetrain and a single soliton of larger amplitude, $A$ = 3.42, moving to the left with velocity $V = -0.16$ (see the middle panel of Fig. 12). As the amplitude of the soliton generated by the breakup of the bi-soliton increases, its velocity changes sign, in agreement with what is shown in Fig. 3 for $\delta = -1$. As the evolution continues, this secondary soliton is eventually overtaken by the initially large-amplitude soliton. Following several subsequent interactions, it gradually loses its energy and is completely destroyed, with its energy being redistributed between the dominant soliton and the surrounding wave field.

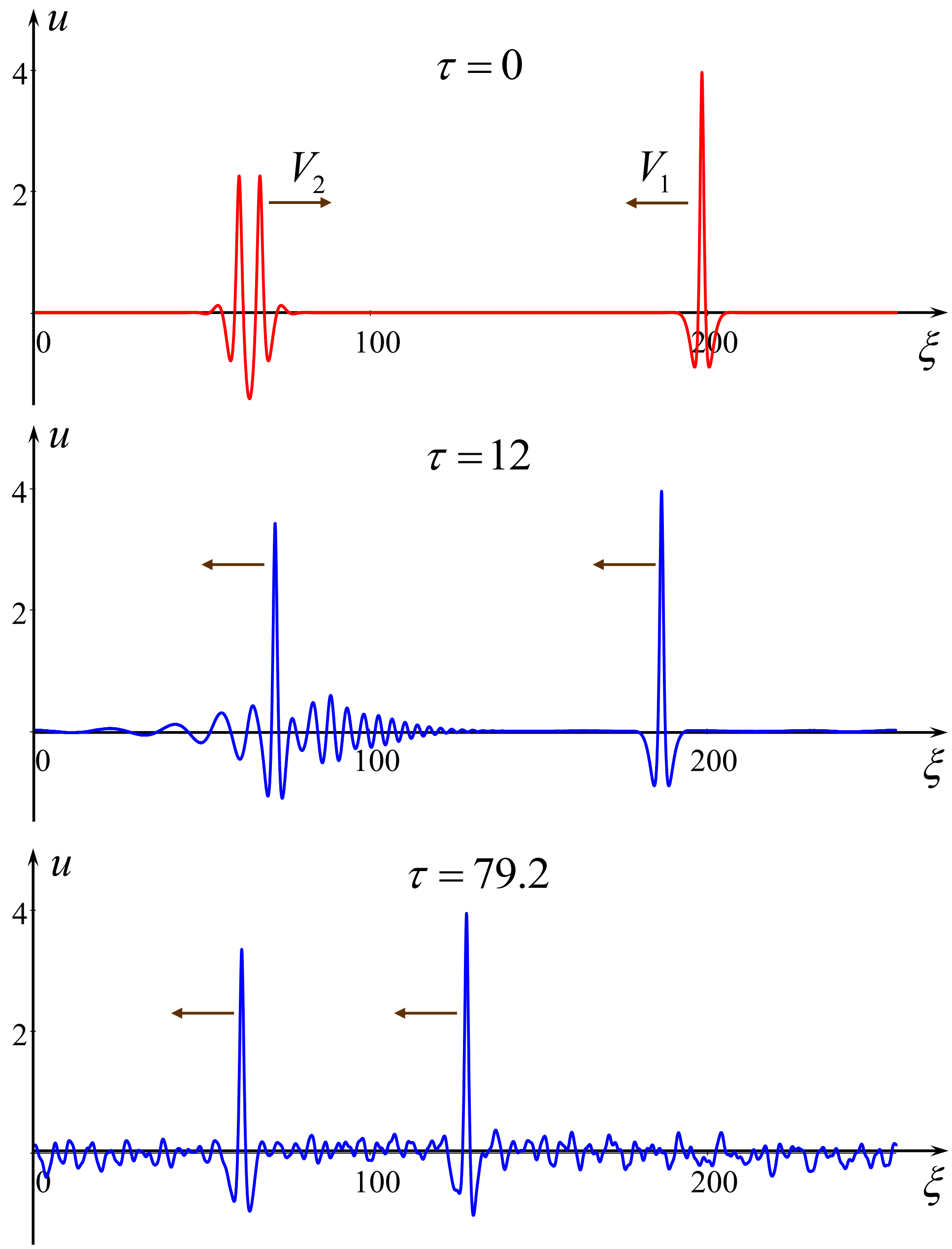

Fig. 12. Head-on initial motion of large-amplitude soliton and unstable bi-soliton with the speeds $V_{1,2} = \pm 1$. The bi-soliton destroys by $\tau = 12$ and forms a single soliton of a large amplitude which moves to the left. After long time, the large amplitude soliton meets the smaller one and terminates it after several interactions in the closed system with periodic boundary conditions.

Another interesting scenario arises when a large-amplitude soliton with $A_1 = 3.954$, propagating to the left with velocity $V_1 = -1$, interacts with a stable bi-soliton of amplitude $A_2 = 2.369$, propagating to the right with velocity $V_2 = 1$. During its propagation, the slightly perturbed bi-soliton exhibits a periodic exchange of amplitude between its constituent solitons: the amplitude of the leading soliton increases slightly while that of the trailing soliton decreases, and then the process reverses without any emission (see Fig. 13). Consequently, the amplitudes oscillate in time, allowing the non-stationary bi-soliton to be interpreted as a particular type of a breather, similar to that observed experimentally in an electromagnetic transmission line by Gorshkov et al. [24].

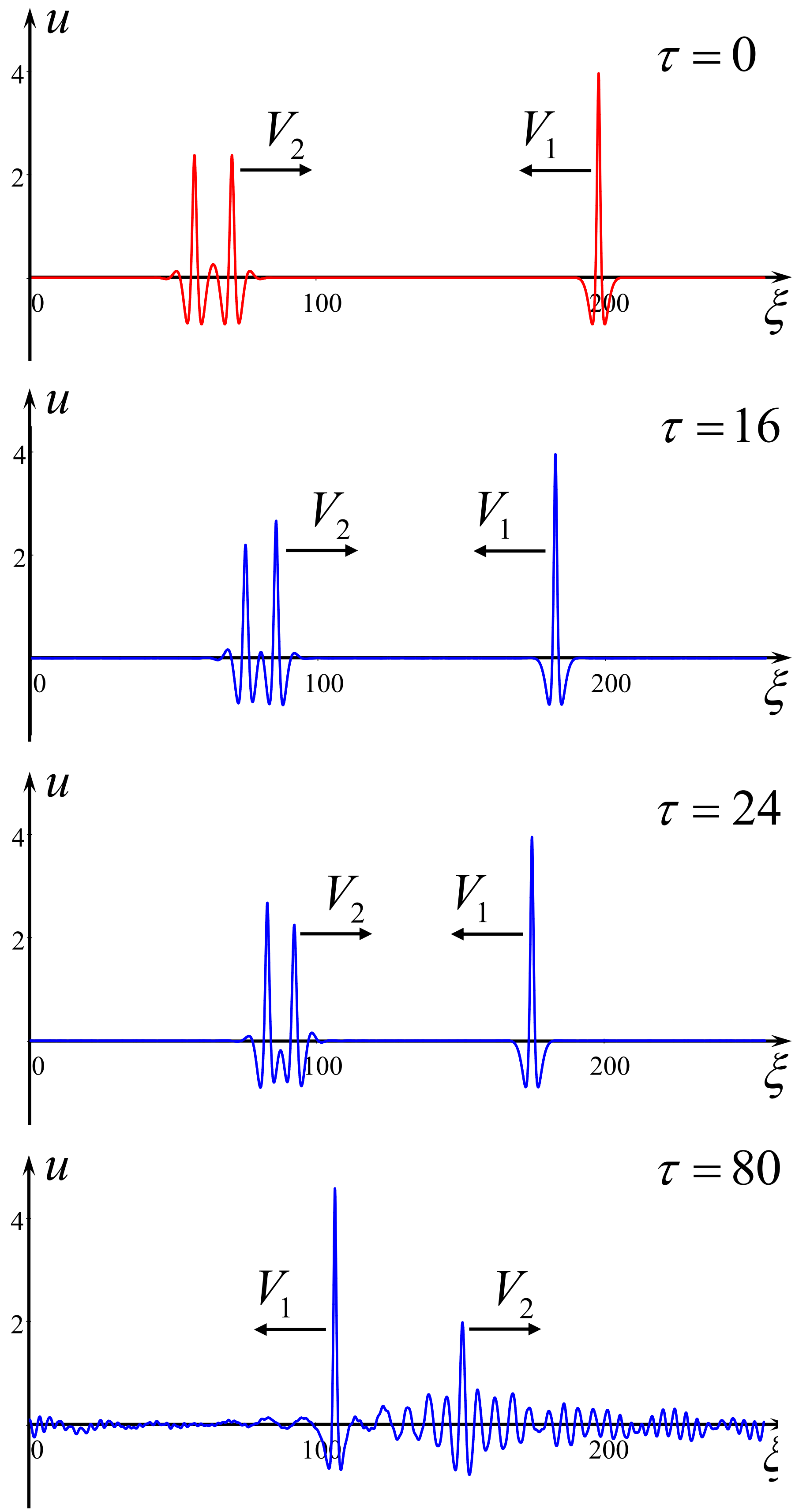

Fig. 13. Head-on initial motion of large-amplitude soliton and a stable bi-soliton with the speeds $V_{1,2} = \pm 1$.

The amplitude exchange within the bi-soliton persists through a few oscillation cycles until the bi-soliton enters into interaction with the large-amplitude soliton. During this interaction, the latter progressively absorbs the bi-solitons, destroying constituent solitons one after another while increasing its own amplitude, as shown in the bottom panel of Fig. 13. As a result, the bi-soliton is completely destroyed, and a dominant soliton-champion emerges with an amplitude 15.5% greater than its initial value (see Fig. 14).

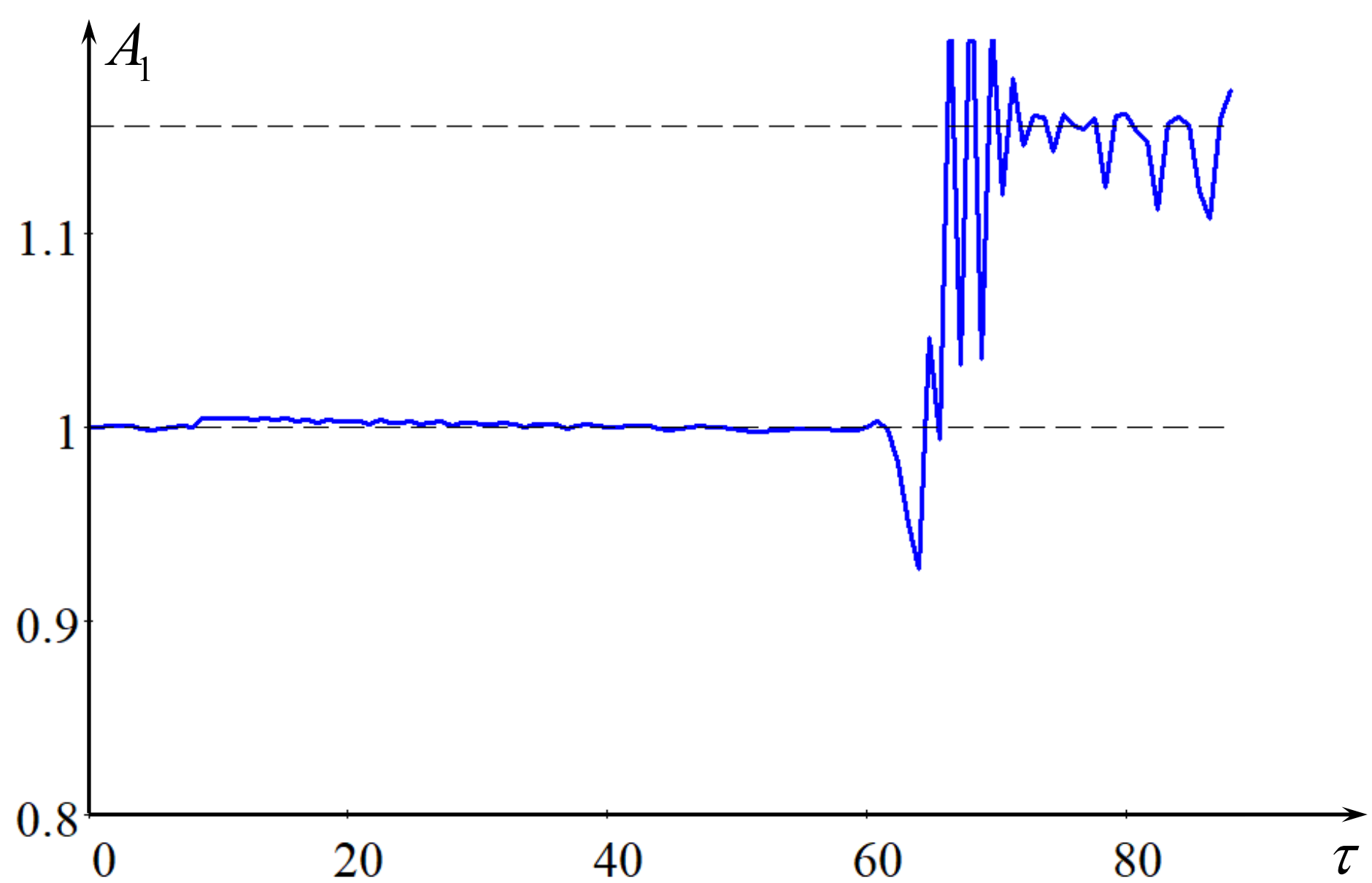

Fig. 14. Time dependence of the largest soliton amplitude in the process of interaction with a stable bi-soliton.

One more example of soliton–bi-soliton interaction is rather interesting and unexpected. We investigated the interaction between a stable large-amplitude bi-soliton ($A_1 = 2.369$, $V_1 = 1$) and a small-amplitude single soliton ($A_2 = 1.769$, $V_2 = 1.5$); see the upper panel of Fig. 15. All solitons moved to the right. During its propagation, the bi-soliton exhibited oscillatory amplitude exchange between its constituent solitons until it encountered the single soliton.

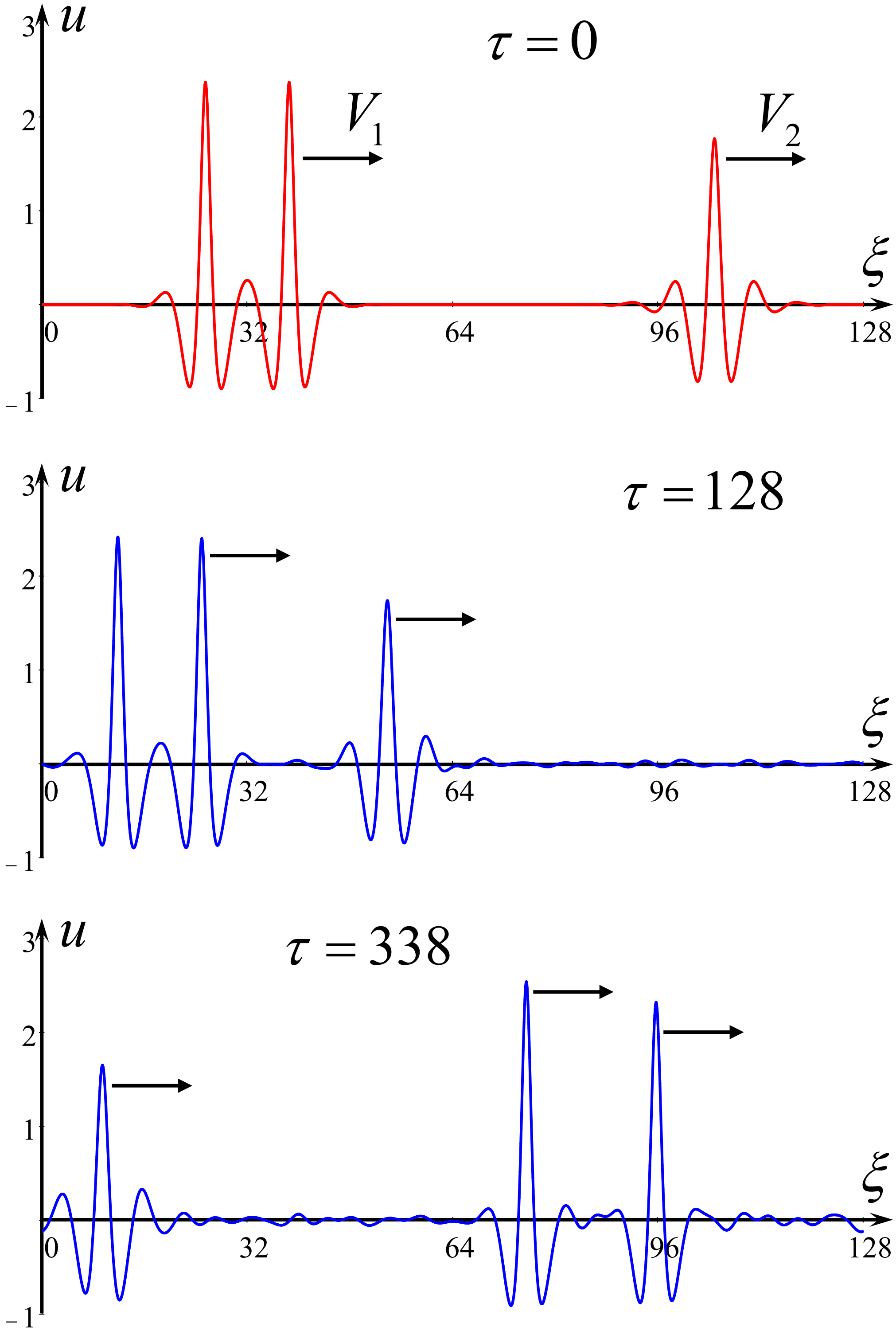

Fig. 15. Interaction of a large-amplitude bi-soliton and small-amplitude single soliton moving with speeds $V_{1,2} = \pm 1$. The bi-soliton splits into two independent solitons after second interaction with the small-amplitude soliton by $\tau = 338$.

Following the first interaction, both wave structures re-emerged with slightly modified amplitudes. The smaller soliton survived the collision, although its amplitude decreased by approximately 1.7%. In addition, a small-amplitude wavetrain was generated during the interaction, as shown in the middle panel of Fig. 15. However, after the second interaction, the bi-soliton split into two independent solitons with slightly different amplitudes; this process is illustrated in the lower panel of Fig. 15. An additional wavetrain was emitted again, and the amplitude of the smallest soliton decreased by a further 2.7%, resulting in a total amplitude reduction of 4.5% after the two interactions.

Figure 16 shows the temporal evolution of the maximum wave-field amplitude. The observed dynamics demonstrate that repeated interactions with a faster-moving single soliton can destabilize an otherwise stable bi-soliton, eventually leading to its breakup into two separate solitons accompanied by the emission of small-amplitude dispersive radiation. Subsequently, three solitons of different amplitudes coexist in the system, with small differences in their propagation velocities. As a consequence, the time intervals between successive collisions become very long, making direct numerical observation of their further interactions computationally demanding.

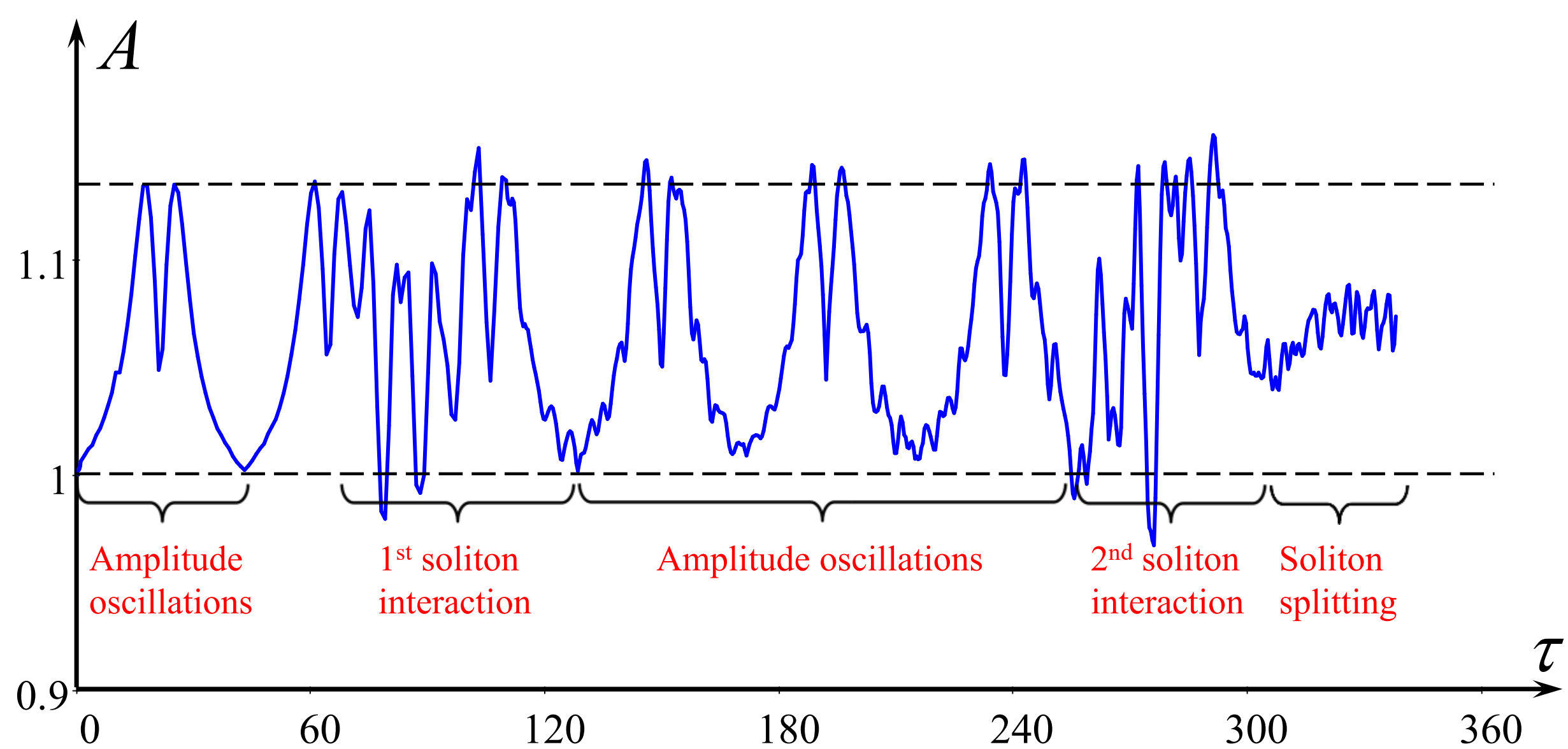

Fig. 16. Time dependence of maximal wave field in the process of interactions of a bi-soliton and small-amplitude single soliton.

In principle, continued evolution of the system would involve multiple collisions among the three solitons, ultimately leading to the emergence of a dominant soliton-champion that absorbs some energy from the smaller solitons and eliminates them. Owing to the extremely long simulation times required, we were unable to perform calculations over a sufficiently extended temporal interval to observe this final stage directly. Nevertheless, based on the interaction scenarios discussed above, the eventual formation of a single dominant soliton appears to be a highly probable outcome.

Thus, the results obtained demonstrate that solitons in the non-integrable GO equation interact inelastically. Such interactions can lead to the complete elimination of a small-amplitude soliton after a single collision if the amplitude of the larger soliton is sufficiently high. However, when the amplitude difference between the interacting solitons is moderate, the smaller-amplitude soliton may survive the initial collision and be gradually eliminated through the subsequent interactions in a closed system with periodic boundary conditions.

As the dominant soliton gains amplitude through these inelastic interactions, its propagation velocity increases accordingly. Consequently, when a large-amplitude soliton encounters a periodic train of smaller-amplitude solitons, it extracts energy from each of them during successive collisions, thereby continuously increasing its own amplitude. As a result, the dominant soliton-champion accelerates on average as it propagates through the soliton train, while the smaller solitons either survive with reduced amplitudes or disappear after the interaction [17].

## 5. Conclusion

In this paper, we have investigated the formation and interaction of solitons in the non-integrable rotation-modified Gardner equation (the Gardner–Ostrovsky equation) with anomalous positive dispersion. We have demonstrated that soliton interactions are inherently inelastic, leading to the emergence of a *soliton-champion*, namely, a single dominant soliton that survives in a closed system with periodic boundary conditions after multiple collisions with smaller-amplitude solitons. We have also shown that pulse-type initial perturbations of sufficiently large amplitude can generate solitons with zero total mass.

In addition, we have examined interactions between solitons and bi-solitons. Our results reveal that bi-solitons can either be completely destroyed through interactions with large-amplitude solitons or split into two independent solitons after repeated interactions with a smaller-amplitude single soliton. In the latter case, the system evolves into a state containing three coexisting solitons with different amplitudes and slightly different velocities.

Although direct numerical simulations over sufficiently long time intervals were not feasible, the observed interaction dynamics suggest that further multiple collisions would eventually lead to the emergence of a single soliton champion, while the two smaller solitons would gradually lose energy and disappear. The long-term evolution of such multi-soliton configurations remains an interesting open problem and deserves further investigation.

**Acknowledgements.** R.F. acknowledges funding received from FAPEMIG, grant No FCT-00208-25.

**Declarations**

**Conflict of interest.** The authors state that there is no conflict of interest.

**Data availability.** All data used in this study are contained in the paper.

**CRediT authorship contribution statement**

R. Fariello: Visualization, Validation, Investigation, Formal analysis.

M. Soares: Visualization, Validation, Investigation, Formal analysis.

Y. Stepanyants: Writing an original draft, Writing review & editing, Visualization, Validation, Supervision, Methodology, Investigation, Formal analysis, Conceptualization.

**R. Fariello**
**Departamento de Ciências da Computação, Universidade Estadual de Montes Claros, Avenida Rui Braga, sn, Vila Mauricéia, Montes Claros, 39401-089, Minas Gerais, Brazil**
**E-mail: ricardo.fariello@unimontes.br**

**M.S. Soares**
**Programa de Pós-Graduação em Modelagem Computacional e Sistemas, Universidade Estadual de Montes Claros, Avenida Rui Braga, sn, Vila Mauricéia, Montes Claros, 39401-089, Minas Gerais, Brazil**
**E-mail: magnusssoares@gmail.com**

**Y.A. Stepanyants**
**School of Mathematics, Physics and Computing, University of Southern Queensland, 487-535 West St., Toowoomba, QLD, 4350, Australia**
**E-mail: Yury.Stepanyants@unisq.edu.au**